\documentclass[12pt,preprint]{aastex}
\usepackage{natbib}
\shorttitle{Dwarf galaxy gas depletion}
\shortauthors{Nichols \& Bland-Hawthorn}

\newcommand{\sol}{\odot}
\newcommand{\HI}{\ifmmode{{\rm H\scriptstyle I}}\else{H${\scriptstyle\rm I}$}\fi}
\newcommand{\HIsub}{\rm{H}\scriptscriptstyle{I}}
\def\gta{\;\lower 0.5ex\hbox{$\buildrel > \over \sim\ $}}
\def\lta{\;\lower 0.5ex\hbox{$\buildrel < \over \sim\ $}}

\begin{document}

\title{Gas depletion in Local Group dwarfs on $\sim$250~kpc scales: Ram pressure stripping assisted by internal heating at early times}
\author{Matthew Nichols}
\email{m.nichols@physics.usyd.edu.au}
\and
\author{Joss Bland-Hawthorn}
\affil{Sydney Institute for Astronomy, School of Physics, The University of Sydney, NSW 2006, Australia}

\begin{abstract}
A recent survey of the Galaxy and M31 reveals that more than $90\%$ of dwarf galaxies within $270$~kpc of their host galaxy are deficient in \HI{} gas.
At such an extreme radius, the coronal halo gas is an order of magnitude too low to remove \HI{} gas through ram-pressure stripping for any reasonable orbit distribution.
However, all dwarfs are known to have an ancient stellar population ($\ga10$~Gyr) from early epochs of vigorous star formation which, through heating of \HI{}, could allow the hot halo to remove this gas.
Our model looks at the evolution of these dwarf galaxies analytically as the host-galaxy dark matter halo and coronal halo gas builds up over cosmic time.
The dwarf galaxies---treated as spherically symmetric, smooth distributions of dark matter and gas---experience early star formation, which sufficiently heats the gas allowing it to be removed easily through tidal stripping by the host galaxy, or ram-pressure stripping by a tenuous hot halo (n$_{\rm H} = 3\times10^{-4}$~cm$^{-3}$ at $50$~kpc).
This model of evolution is able to explain the observed radial distribution of gas-deficient and gas-rich dwarfs around the Galaxy and M31 if the dwarfs fell in at high redshifts ($z\sim3$--$10$).

\end{abstract}

\section{Introduction}
In recent years, the Sloan Digital Sky Survey \citep{York2000} has more than doubled the number of known dwarf galaxies 
that orbit the Galaxy \citep{Willman2005,Willman2005a,Zucker2006,Belokurov2006,Zucker2006a,Belokurov2007,Irwin2007,Walsh2007,Belokurov2010,Willman2010}.
These newly discovered dwarf galaxies, being among the most dark-matter dominated structures in the universe, help address the perceived imbalance between the number of predicted low mass dark matter substructure and those observed today, the ``Missing Satellites Problem'' \citep{Klypin1999,Moore1999}.
As low mass substructures are the building blocks upon which galaxies are formed under the $\Lambda$CDM paradigm, the ability to probe dwarf galaxies provides invaluable knowledge into the history and formation of the Local Group \citep{Tolstoy2009,Karlsson2011}.

Although the dark matter component can be tracked through the use of high resolution {\em N}-body simulations \citep{Diemand2007,Diemand2008,Springel2008,Wetzel2010}, a complete model of the evolution of these dwarfs is still elusive.
All of the known dwarfs show signs of old stellar populations \citep{Tolstoy2009} with more of the star formation occurring in discrete bursty periods of star formation---of order $25\%$ of total star formation \citep{Lee2009}---compared to more massive galaxies which experience roughly continuous star formation.
This bursty behaviour, with bursts separated by gigayears, does not necessarily require interactions to trigger star formation \citep{Brosch2004} with the blow-out and subsequent infall of neutral gas sufficient to create bursts in isolated dwarfs \citep{Valcke2008,Quillen2008}.

In a recent development, semi-analytic models of galaxy formation, combined with results from high-resolution dark matter simulations, have been used to model the physical properties of a Galactic dwarf galaxy system \citep{Li2010}. These models however do not yet account for the underabundance of \HI{} detected in these dwarf galaxies over the last $40$~years \citep[hereafter GP09]{Einasto1974, Grcevich2009}.
The dependence of the \HI\ deficiency on galactocentric radius is evidence of significant tidal and/or ram-pressure stripping of the dwarf galaxies which GP09 attribute towards close-in, potentially highly eccentric orbits, allowing the \HI\ to be removed by a hot halo surrounding the Galaxy.
This removal for close pericentres ($\la50$~kpc) has been explained by a combination of tidal and ram pressure forces \citep{Mayer2006}.

But we now show that {\it unassisted} stripping fails by an order of magnitude to explain the phenomenon of dwarf galaxy depletion, on scales of $\sim250$~kpc as observed today, for any reasonable orbit families (\S\ref{sec:Mod}).
In \S\ref{sec:Res}, we include the effects of early star formation and find that feedback-assisted stripping is essential to explain the observed effect. Using models of dwarf infall, we then compare results from our model to the observed properties of the \HI{} abundance in dwarfs today in \S\ref{sec:Comp}.
Finally, a discussion of the limitations of this model and what impact that will have on the results is given in \S\ref{sec:Conc}.

\section{Dwarf Galaxy Model}\label{sec:Mod}
A dwarf galaxy with an eccentric orbit will experience a wide range of extremes as it travels along its orbit, from a low velocity through sparse gas at its apogalacticon to a large velocity through dense gas with a strong external radiation field at its perigalacticon.
All these environments must be modelled.
We describe this wide range of environments by considering four scenarios: the orbital path of the dwarf around the host galaxy, discussed in \S\ref{ssec:orbit}; the heating and cooling of the gas present, discussed in \S\ref{ssec:heating}; how tidal or ram pressure stripping removes gas from the dwarf in \S\ref{ssec:RPS}; and how this stripping can be influenced by stellar formation and feedback, detailed in \S\ref{ssec:starform}.
All four, to varying degrees, will be influenced by how much gas the dwarf galaxy has at the beginning of its orbit and the properties of the dark matter halo that envelopes the dwarf.

The profile of the dwarf galaxy dark-matter halo is not a simple choice, with no solid consensus on which profile most accurately describes the low-mass subhalos present in dwarf galaxies. 
Observationally, the dark-matter profile of dwarfs cannot be distinguished between a cuspy profile and that of a cored, constant-density profile \citep{Walker2009}.
Both of these profiles also arise through simulations with the cuspy Einasto profile arising through large {\it N}-body, dark-matter only, simulations \citep{Springel2008}.
With the addition of baryons, it has been argued that an even steeper central core---through the process of baryonic
contraction---can be present than predicted by the dissipationless $N$-body simulations \citep{Blumenthal1986,DelPopolo2009,Napolitano2010}.
However, by coupling the baryons with the dark matter, and allowing dynamical and angular momentum transfer to occur, constant density cores with Burkert like profiles \citep{Burkert1995} are able to develop in simulation \citep{El-Zant2001,DelPopolo2009}.
Since our major focus is on dark matter-halos with low gas fraction, whereby both baryon--dark-matter coupling and baryonic contraction will have a smaller impact, we ignore these competing factors and use the Einasto profile.

The choice of profile for the host galaxy is less contentious, and is chosen to have an evolving Einasto profile.\
Although the host contains more baryons than the dwarf galaxies we are interested in, due to its much larger mass baryonic contraction processes have a lower net effect.
Thus the evolution of the host galaxy is dominated by the infall of dwarfs onto the galaxy.
The evolution of the host of this galaxy is assumed to be similar to that of the Milky Way, with a growth rate described by \citet{McBride2009} leading to a final virial mass of $1.37\times10^{12}$~M$_\odot$ \citep[for an Einasto model with the same circular velocity at virial radius as the NFW model in][]{Smith2007}, to give a mass at redshift $z$ of
\begin{equation}
  M(z) = \left[1.69\times10^{-19}\int^0_z (1+1.17z)\sqrt{\Omega_0(1+z^3)+\Omega_\Lambda}\frac{\mathrm{d}t}{\mathrm{d}z}\mathrm{d}z + 0.96\right]^{-7.87}10^{12}~\mathrm{M}_\odot, \label{eq:Mz}
\end{equation}

We assume that any cold gas is of a constant density, and gas in the warm and hot phases in the dwarf is in hydrostatic equilibrium with the density of the gas then given by
\begin{equation}
  n(r) = n(0)\exp[-V(r)/c_s^2],
\end{equation}
where $n(0)$ is the central density, $c_s$ is the speed of sound in the gas and $V(r)$ is the Einasto potential given by \citep{Cardone2005,Nichols2009}
\begin{equation}
  V(x)=\frac{3}{4} v_s(t)^28^{-1/\alpha}\exp(2/\alpha)\alpha^{-1+3/\alpha}\left[2^{1/\alpha}\alpha^{-1/\alpha}\gamma(2/\alpha,2x^{\alpha}/\alpha)-\gamma(3/\alpha,2x^{\alpha}/\alpha)/x-1\right],
\end{equation}
with $x\equiv{}r/r_s(t)$ the scaled radius, $r_s(t)$ the halo scale radius, $v_s(t)$ the characteristic scale velocity of the dark matter halo, $\alpha$ the Einasto free parameter, and $\gamma(a,x)$ the lower incomplete gamma function.

The central density of the gas is calculated by assuming the calculated central density of Leo T in GP09 of $n_{\rm H}=0.45$~cm$^{-3}$ is comprised of gas in pressure equilibrium with a mass ratio $1:3$ of cold gas to warm gas, matching the total ratio of masses within Leo T from which this value was calculated \citep{Ryan-Weber2008}.
This gives a warm gas density of $0.05$~cm$^{-3}$ and cold gas density of $1.5$~cm$^{-3}$.
The hot phase central density is assumed to be in pressure equilibrium with the cold and warm phases.

\subsection{Orbit Of The Dwarf}\label{ssec:orbit}
The orbit of the dwarf galaxy within the larger host galaxy is given by the shape of the potential.
As the host galaxy evolves, the halo scale radius and scale velocity will change according to \citep{Sternberg2002}
\begin{eqnarray}
  r_s(t) &=& \left(\frac{3}{4\pi{}\Delta(t)\rho_u(t)}\right)^{1/3}\frac{M_{\rm vir}(t)^{1/3}}{x_{\rm vir}(t)},\\
  v_s(t) &=& \left(\frac{4\pi}{3G^3\Delta(t)\rho_u(t)}\right)^{1/6}M_{\rm vir}^{1/3}\sqrt{\frac{x_{\rm vir}(t)}{f_m(x_{\rm vir}(t))}},
\end{eqnarray}
where $\Delta$ is the characteristic overdensity of dark matter halos, $\rho_u$ is the mean baryonic density of the universe, $x_{\rm vir}\equiv{}R_{\rm vir}/r_s$ is the concentration parameter and $f_m$ is the mass profile as given in \citet{Nichols2009}.

The characteristic overdensity is given by \citep{Bryan1998}
\begin{equation}
  \Delta = \frac{18\pi^2+82(\Omega(z)-1)-39(\Omega(z)-1)^2}{\Omega(z)},
\end{equation}
and the mean baryonic density by
\begin{equation}
  \rho_u = \frac{3H(z)}{8\pi{}G}\Omega(z).
\end{equation}

For an Einasto profile, the concentration parameter, $x_{\rm vir}$, also referred to as $c$, is \citep{Duffy2008}
\begin{equation}  x_{\rm vir} = 8.82(1+z)^{0.87}\left(\frac{M_{\rm vir}}{2\times10^{12}~h~\mathrm{M}_\odot}\right)^{-0.106}
\end{equation}

In this growing potential there is no angular dependence and therefore the angular momentum is conserved; such that
\begin{equation}
  L^2 = \frac{3\cdot2^{-1-3/\alpha}\exp(2/\alpha)v_s^2(t)\alpha^{-1+3/\alpha}\int^{r_s/r_a}_{r_s/r_p}\gamma(3/\alpha,2x'^{-\alpha}/\alpha) dx'}{1/r_a^2 - 1/r_p^2},
\end{equation}
where $L^2$ is the per unit mass angular momentum squared and $r_p$ and $r_a$ are the peri and apogalacticon radii that would occur if the potential was static.

We present our results in terms of both orbit circularity and eccentricity in recognition of the fact that both are in common use when discussing CDM model.
We define the eccentricity of the orbit in terms of the ratio of the apogalacticon to the perigalacticon, $\epsilon=(r_a-r_p)/(r_a+r_p)$, and the corresponding circularity to be $\eta=\sqrt{1-\epsilon^2}$.
These parameters are not conserved throughout the orbit due to the growing potential, and when referred to we use the final state of these parameters.

We also add dynamical friction as a non-conservative force, with a magnitude of the instantaneous dynamical friction timescale multiplied by the velocity \citep{Zhao2004}
\begin{equation}
  t_{\rm fric}^{-1} = \frac{4\pi{}G^2\rho{}M_{\rm dw}}{v_{\rm circ}^3}\left(\frac{2.5}{4/3 + (v/v_{\rm circ})^3}\right),
\end{equation}
giving the equations of motion as
\begin{eqnarray}
  \ddot{r} &=& r\dot{\theta}^2 - \frac{\partial{}V(r,t)}{\partial{}r} - \frac{4G\pi{}r^3\rho_u(z)}{r^2} - \dot{r}t_{\rm fric}^{-1}, \\
  \ddot{\theta} &=& -2\frac{\dot{r}\dot{\theta}}{r} - \frac{\dot{\theta}}{r}t_{\rm fric}^{-1},
\end{eqnarray}
with the additional term in the radial direction arising from the background contribution to the density of the dark matter, which becomes non-negligible beyond $r\approx{}2r_{\rm vir}$.

\subsection{Heating And Cooling}\label{ssec:heating}

Throughout the passage of the orbit, the gas present within a dwarf galaxy will be exposed to a variety of radiation fields, with the most powerful being the radiation field from the host galaxy, the extragalactic background light and the stellar radiation field produced from high mass, short lived stars formed within the dwarf galaxy.
To examine if the changing orbit itself is enough to produce the observed galactocentric effects, we first examine the system without the internal stellar radiation field.

Without internal star formation, two contributions to the radiation field will dominate the heating and ionization of the dwarf galaxy, namely the extragalactic background field and the radiation field from the host galaxy which will dominate when the dwarf is within $\sim100$~kpc of the host galaxy's centre.
For the extragalactic background field we use the redshift-dependent fields of \citet{Faucher-Giguere2009} and assume that all the radiation enters the dwarf galaxy radially. To a good approximation,
the radiation field of the host $L_\star$ galaxy is assumed to be a delta function at $13.6$~eV with the number of photons impacting the dwarf galaxy at a distance $r$ from the disk given by \citep{JBH1999}
\begin{equation}
  \varphi(r,z) = \frac{0.01\times2.6\times10^{53}}{r^2}\frac{{\rm SFR}(z)}{{\rm SFR}(0)},
\end{equation}
with the $0.01$ factor arising from a $6\%$ escape fraction vertically from the disk and isotropized over the halo.
SFR$(z)$ is the star formation rate at redshift $z$, given by \citep{Just2010}
\begin{equation}
  {\rm SFR}(t) = 2\frac{(t+1.13~\mathrm{Gyr})(11.7~\mathrm{Gyr})^3}{(t^2+[7.8~\mathrm{Gyr}]^2)^2}~\mathrm{M}_\odot~\mathrm{yr}^{-1},
\end{equation}
where $t$ is the time since disk formation, which we assume formed $12$~Gyr before the present day.

These two radiation fields combine to heat and ionize the cold gas that is present in the dwarf galaxy.
The gas is ionized according to the absorbed radiation field, for which a radial radiation field yields
\begin{equation}
  \xi = \frac{2R_{\rm dw}^2}{\hbar}\int_{\log(13.6~\mathrm{eV}/[2\pi\hbar])}^{\log(7.25\times10^{17})} J(r,\nu,z)(1-\exp[-N_{\HIsub}\sigma(\nu)]) \mathrm{d}\log\nu,
\end{equation}
with $R_{\rm dw}$ the radial size of the cold gas in the dwarf and $J(r,\nu,z)$ is the combined radiation field of the Galaxy and the extragalactic UV field at a distance $r$, frequency $\nu$ and redshift $z$ in ergs~s$^{-1}$~cm$^{-2}$~Hz$^{-1}$.

These radiation fields will also heat the gas, with the heating (considering only H and He) given by \citep{Wolfire1995b}
\begin{equation}
  \Gamma_{XR} = \frac{2R_{\rm dw}^2}{\hbar}\int_{\log(13.6~\mathrm{eV}/[2\pi\hbar])}^{\log(7.254\times10^{17})} J(r,\nu,z)(1-\exp[-N_{\HIsub}\sigma(\nu)])E_h(E,n_{\rm e}/n_{\rm H}) \mathrm{d} \log\nu,
\end{equation}
where $E_h$ is the heating per primary electron given by \citet{Wolfire1995b}.

The cooling, $\Lambda(T,Z)$, is calculated by metal line cooling at a metallicity of $0.1Z_\odot$ with the fits provided by \citet{Schure2009} and He collision cooling by \citet{Dalgarno1972}.

As more particles are ionized than can be warmed, any excess ionizations (after recombination) are converted to heating with all $13.6$~eV going to heat the gas.
The transfer from cold neutral gas to warm ionized gas, is then given by
\begin{eqnarray}
  \dot{M}_{\rm XR} &=& \frac{\Gamma_{\rm XR}}{3/2k_b(T_w-T_c)m_{\rm H}},\nonumber\\
  \dot{M}_{\rm cool} &=& \frac{8}{3}\pi{}m_{\rm H}r_{s}^3\frac{\Lambda(T,z)}{k_{\rm b}T}n_{{\rm H},0,w}^2\int^{x_{{\rm edge},w}}_0 x^2f_{\rm gas}^2(x)\mathrm{d}x,\nonumber\\
  \xi_{\rm net} &=& \xi-4\pi{}n_{{\rm H},0,w}^{2}2\times10^{-10}T_{w}^{-0.75}\int^{x_{{\rm edge},w}}_0 x^2f_{\rm gas}^2(x)\mathrm{d}x,\nonumber\\
  \dot{M}_{\rm cold} &=& \dot{M}_{\rm cool} - \dot{M}_{\rm XR} - \frac{\xi_{\rm net}-\dot{M}_{\rm cool}+\dot{M}_{\rm XR}}{1.5k_b(T_w-T_c)m_H},
\end{eqnarray}
where $T_c$ and $T_w$ are the temperatures of the cold and warm phase respectively, $n_{{\rm H},0,w}$ and $x_{{\rm edge},w}$ is the central density and scaled radii of the edge of the warm phase respectively, and $f_{\rm gas}$ is the Einasto gas distribution function.
Correspondingly $\dot{M}_{\rm warm} = - \dot{M}_{\rm cold}$.

Without any other source of heating (see \S\ref{ssec:starform}), the cooling prevents any significant heating in the model, keeping most gas as cold gas with only a small amount of warm gas in the centre. 

\subsection{Tidal and Ram Pressure Stripping}\label{ssec:RPS}

Gas in the warm or cold phase may be stripped through a combination of tidal forces and that of ram pressure arising from the orbit through the hot halo of the host galaxy.

The tidal forces will not only potentially remove gas from the dwarf galaxy but will strip down the dark matter from the halo that surrounds the dwarf.
The tidal extent of the dwarf's halo, neglecting the contribution of baryonic mass, is given by \citep{Hayashi2003}
\begin{equation}
  \frac{m(R_t)}{R_t^3} = \left[2-\frac{r}{M(r)}\frac{\partial{}M}{\partial{}r}\right]\frac{M(r)}{r^3},
\end{equation}
where $R_t$ is the tidal radius of the dwarf, and $r$ is the orbital radius of the dwarf from the host galaxies centre.
For an Einasto profile, this tends not to have an analytic solution, and for reasons of speed is only solved initially, before using the derivative to calculate the change in tidal radius over time.
Any gas that extends above the tidal radius is considered to be no longer bound, and hence lost, in order to simplify calculation.
A more extensive treatment involving re-accretion onto dwarfs requires hydrodynamical treatments that are now under way.
Within our models, tidal stripping does not produce significant gas loss with Galactic and extragalactic ionizing fields keeping most gas deep within the tidal radius.

Working at the same time as the tidal stripping is ram pressure stripping.
Here we use the \citet{McCarthy2008} formalism whereby the ram pressure removes shells of gas from the dwarf galaxy, assuming that the gas distribution behaves as an isothermal sphere beyond the cloud edge if
\begin{equation}
  \rho_{\rm gal}(r)v^2 > \frac{\pi{}}{2}\frac{GM_{\rm dw}(R)\rho_{\rm gas}(R)}{R}.
\end{equation}
This formalism, originally used for a hot halo, can be carried over to the warm and cold phases by assuming spherical symmetry and a $\rho\propto{}r^{-2}$ density drop off outside the cloud edge.

We note that when using the realistic approximation $\sigma\sim{}v_{\rm circ}/\sqrt{2}$, this is $3\pi{}$ times more difficult to strip than the \citet{Gunn1972} equation used by GP09, i.e.
\begin{equation}
  n_{\rm gal} \sim \frac{\sigma^2n_{\rm gas}}{3v^2_{sat}}.
\end{equation}

If the stripping does occur, then the rate of stripping will be dependent upon the speed of the shock induced in the gas \citep{Mori2000}, such that
\begin{equation}
  v_{fs} = \frac{4}{3}\sqrt{\frac{\rho_{\rm gal}(r)}{\rho_{\rm gas}(R)}}v .
\end{equation}
Here we assume that the halo stays spherically symmetric at all times, and hence the rate of the change in radius will be half the speed of the shock, i.e. $\dot{R} = 0.5v_{fs}$.

The density of the host galaxy halo will be dependent upon the temperature of the halo.
We use an isothermal halo with the temperature set at the virial temperature of the halo at any given redshift.
We set an unchanging central density to give a density of $3\times10^{-4}$~cm$^{-3}$ at $50$~kpc at the 
present day.
These values are consistent with isothermal halo parameters from \citet{Battaglia2005}, the values used by \citet{JBH2007} (which uses $T=1.75\times10^6$~K, $n_{\rm H}=2\times10^{-4}$~cm$^{-3}$ at $55$~kpc), and the high-entropy Galactic halo model of \citet{Kaufmann2009}. 

For this stripping, the phases are assumed to be independent, and the cold phase can be removed at the same time as the warm phase if the ram pressure is large enough.
The rate of change of mass is then simply
\begin{equation}
  \dot{M} = 4\pi{}r^2\rho_{\rm gas}(R)\dot{R}.
\end{equation}

With only Galactic and extragalactic ionizing fields, ram-pressure, like tidal stripping, does not produce large changes in gas mass even over cosmic time.
Even with more favourable conditions to ram pressure stripping, such as the Gunn-Gott criterion and a lower temperature halo (resulting in a denser inner portion), only a small amount of gas is lost---less than $10^5$~M$_\odot$---corresponding to the cold gas heated by the Galactic and extragalactic UV fields to a warm phase and then lost to ram-pressure stripping.

These results are in conflict with much more substantial gas loss has been measured in hydrodynamical simulations with ram pressure alone able to strip the gas from close in ($r_{\rm peri} < 50$~kpc), low mass halos \citep{Mayer2006}.
However, at larger perigalacticons, dwarfs in \citet{Mayer2006} are still able to retain much of their centrally concentrated gas, suggesting another form of heating is required to allow the removal of gas from dwarfs with larger perigalacticons \citep[e.g. the perigalacticon of Sextans and Draco is expected to be $>50$~kpc][]{Lux2010}.

\subsection{Star Formation and Feedback}\label{ssec:starform}
A dwarf consisting of cold, dense gas is resistant to tidal and ram-pressure stripping with only a thin skin, ionized by the Galactic of extragalactic UV fields being removed.

Early star formation however, will heat this cold gas, raising it in the potential well and making it more easily stripped.
This warm ionized gas, at the same pressure as cold gas will occupy $\sim60$ times the volume, and being much less dense will become even easier to strip than its height in the potential well would indicate.



The star formation in dwarfs is considered to consist of periods of low-level star formation, during which short bursts are induced which increase the star formation by a factor of $3$, consistent with dwarfs observed by \citet{Lee2009}.
We assume these bursts are triggered by perigalacticon passages, induced by shocks created through tidal interactions \citep{Pasetto2010}.
There is also evidence that bursts may be triggered by the re-accretion of heated and expanded gas \citep{Valcke2008}, however, much of this gas will be stripped while in the warm phase and far from the centre preventing the re-accretion and subsequent starburst from occurring.

The base star formation rate is taken to be similar to the dwarfs that surround M31, with a star formation rate \citep{Kaisin2006}
\begin{equation}
  \log (\mathrm{SFR}~{\rm M}_\sol~{\rm yr}^{-1})  = 1.4\log M_{\HIsub}+k,
\end{equation}
where $k$ is a constant calculated by assuming that the gas will be completely depleted at a time $t=5/H_0\sim70~$Gyr.
For an initial gas mass of $5\times10^7$~M$_\odot$, $k=-13.62$.

The UV--X-ray spectrum produced by the star formation in the dwarfs---which is responsible for the majority of heating---was calculated with {\tt Starburst99} \citep{Leitherer1999}; when a burst occured the change in the spectrum was calculated by summing many short bursts together to produce an approximately continuous change in the spectrum.
Above the Lyman limit, this spectrum (in ergs~$s^{-1}$~s$^{-1}$~cm$^{-2}$~Hz$^{-1}$~sr$^{-1}$) and its changes during a burst was well fitted by

\begin{equation}
  \log J_\nu = 2+J_{i}+\log[{\rm SFR~M}_\sol~{\rm yr}^{-1}]-3.54\log \nu ,
\end{equation}
where $J_{i}$ changes linearly over a $5$~Myr period at the beginning and end of each burst from $J_{i}=96.15$~(ergs~$s^{-1}$~s$^{-1}$~cm$^{-2}$~Hz$^{-1}$~sr$^{-1}$) at the low level continuous star formation rate to $J_{i}=96.63$~(ergs~$s^{-1}$~s$^{-1}$~cm$^{-2}$~Hz$^{-1}$~sr$^{-1}$) during a burst.

The addition of star formation will also result in supernova superheating the gas well above the temperature of the warm phase
Any power from the supernova is assumed to go into heating gas into a hot ($10^6$~K) phase.
In reality, the energy from the supernova will also go into expanding the cold and warm phases, making them more susceptible to stripping, but due to the limitations of a smooth density profile we assume this energy is instead used to heat the gas.

We find the power output of a supernova for continuous, non-bursty star formation is given by
\begin{equation}
  \log[P_{\rm SN}~{\rm erg~s}^{-1}] = \log[{\rm SFR~M}_\sol~{\rm yr}^{-1}] + 40.0.
\end{equation}

After a burst has commenced, the power output by the supernovae increases over $38$~Myr, with the power over this period of increasing output given by
\begin{equation}
  \log[P_{\rm SN}~{\rm erg~s}^{-1}] = \log[{\rm SFR~M}_\sol~{\rm yr}^{-1}] + 2 + 25.6\tan^{-1}[24(0.1t+1.7)],
\end{equation}
where $t$ is the time in Myr since the burst commenced.

During the increased star formation associated with the bursty period \citep[lasting $\le10^7$~yr][]{Sharp2010} the power output is
\begin{equation}
  \log[P_{\rm SN}~{\rm erg~s}^{-1}] = \log[{\rm SFR~M}_\sol~{\rm yr}^{-1}] + 42.0.
\end{equation}

After the burst has ended and star formation drops back down to the continuous star formation rate, the power output subsides as massive stars die off.
Over a period of $40$~Myr, the power output declines according to
\begin{equation}
  \log[P_{\rm SN}~{\rm erg~s}^{-1}] = \log[{\rm SFR~M}_\sol~{\rm yr}^{-1}] + 42.0 - (0.0125t') \label{eq:PSN},
\end{equation}
where $t'$ is the time since the starburst has ended in Myr.


In this simplified model, we assume that any power from the supernova is dissipated in the gas and goes into heating the cold and warm gas at $100\%$ efficiency and split between the cold and warm phase in proportion to their mass.
The mass loss rate will then be the power dissipated into that phase divided between the energy difference between that phase and the hot phase.
\begin{eqnarray}
  \dot{M}_{c, {\rm SN}} = \frac{P_{\rm SN}m_{\rm H}}{1.5k_b[(T_h-T_c) + (T_h-T_w)\frac{M_{\rm warm}}{M_{\rm cold}}]},\nonumber\\
  \dot{M}_{w, {\rm SN}} = \frac{P_{\rm SN}m_{\rm H}}{1.5k_b[(T_h-T_w) + (T_h-T_c)\frac{M_{\rm cold}}{M_{\rm warm}}]}.
\end{eqnarray}
Although the efficiency of supernova heating is typically much lower, there is only a small amount of material that is heated into the hot phase, with much more material being lost through the warm phase.
In this limit, the snowplough phase of supernova expansion, where a large portion of energy goes, may have a noticeable contribution by raising gas outside of the potential well and we keep the efficiency at unity to partially compensate for this effect.

\begin{figure}
  \begin{center}
    \includegraphics[angle=-90,width=\textwidth]{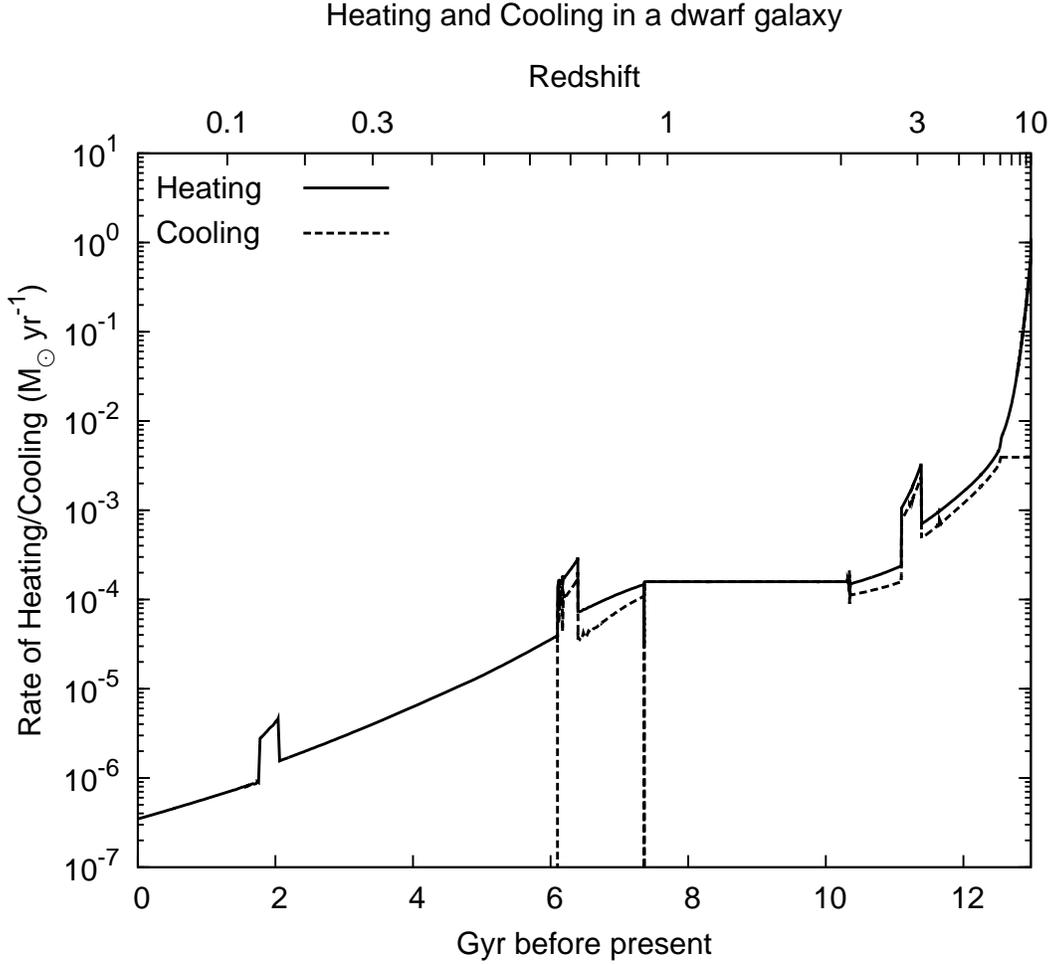}
    \caption{The relative rates of cooling for a dwarf galaxy finishing with perigalacticon of $80$~kpc and circularity of $0.9$ (eccentricity 0.44). Initially the cooling is minimised by warm gas being forced outside the tidal radius of the dwarf, preventing it recooling onto the dwarf, after the first pericentre passage, the cooling and heating reach equilibrium, until the next pericentre, when the warm gas drops below $10$~M$_\sol$ and is no longer tracked. The drop in cooling around $7.5$~Gyr is an artifact of the method of integration when heating and cooling are no longer in equilibrium and does not change any results.}\label{fig:heatcool}
  \end{center}
\end{figure}

Under these conditions the internal star formation greatly exceeds the extragalactic UV background at early times, making it the dominant source of heating.
The cooling of gas from the warm phase to the cold phase can approach the level of UV--X-ray heating, however, the amount of warm gas that is able to be held on by the halo is limited, preventing most gas from recooling.
The relative rates of heating by the combined radiation fields (internal star formation, Galactic and extragalactic) and the cooling rate is shown in Fig. \ref{fig:heatcool} for a model with perigalacticon of $80$~kpc and circularity of $0.9$ (eccentricity of $0.44$).
Very little gas is moved to the hot phase from supernova bursts, much more gas is heated to the warm phase which quickly extends beyond the tidal radius, resulting in large amounts of gas loss from tidal stripping.
Only in more massive dwarfs with correspondingly larger central densities will sufficient quantities of warm gas be able to cool back to a cold phase to allow for significant supernova heating. 

\subsection{Method and Initial Conditions}\label{ssec:IC}
The gas and dark matter masses were tracked by solving the time dependent equations---equations (\ref{eq:Mz})--(\ref{eq:PSN})---with the {\tt odeint} routine \citep{Press1992} and encoded in {\tt python}.
Once a gas phase dropped below $10$~M$_\odot$, it was no longer tracked and assumed to no longer have any gas in that phase.

The dwarfs were initiated with $5\times10^7$~M$_\sol$ for the cold phase and $100$~M$_\sol$ for the warm and hot phases.
The dark matter halo of each dwarf galaxy was set with a virial mass of $1\times10^{9}$~M$_\sol$ for models beginning at $z=1$ and $z=3$ and a virial mass of $3\times10^{8}$~M$_\sol$ for $z=10$.
For all starting points, the predicted mass within $300$~pc is about $10^{7}$~M$_\sol$, consistent with the mass inferred for dwarf galaxies \citep{Strigari2008}.

\begin{figure}
  \centering
  \includegraphics[angle=-90,width=\textwidth]{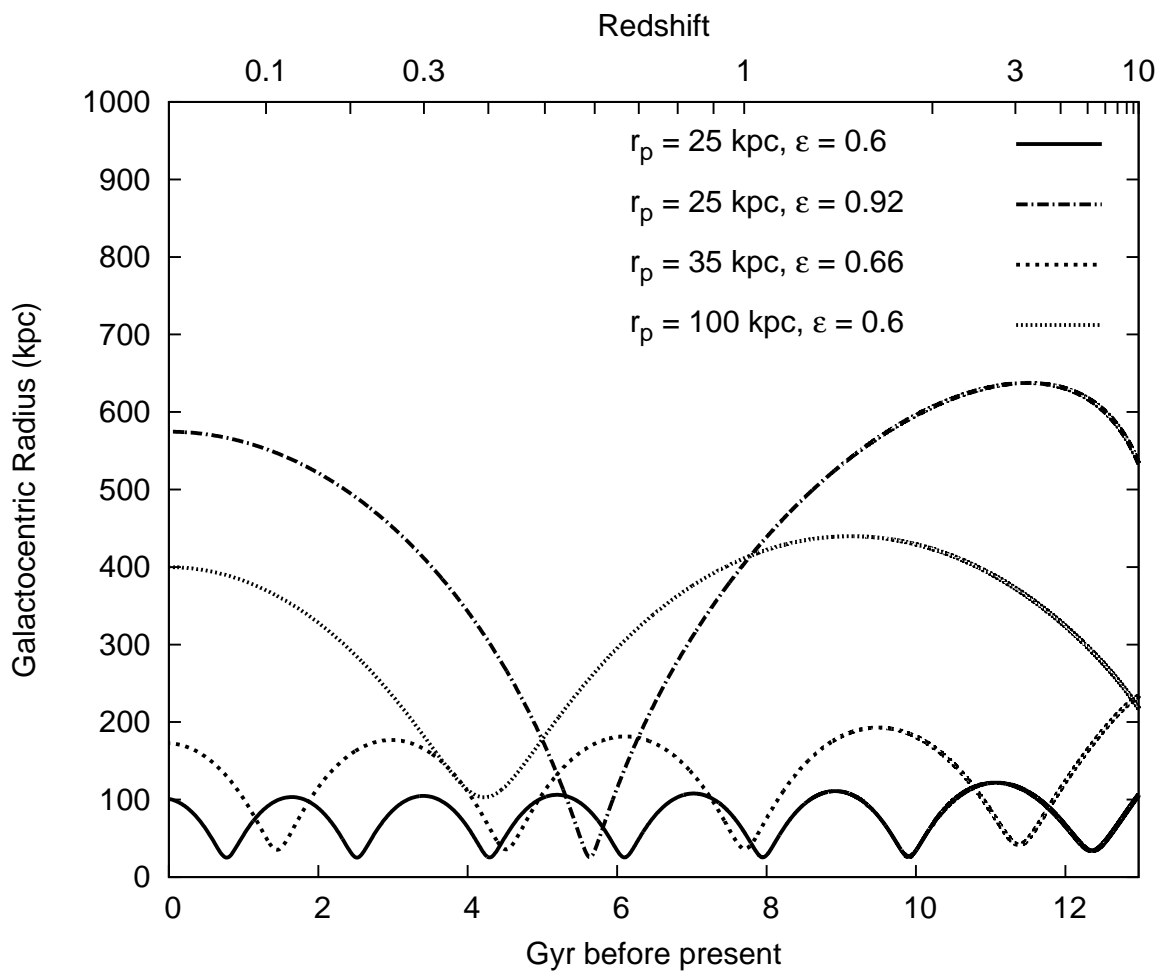}
  \caption{A small subset of orbits of dwarf galaxies of mass $M_{\rm vir}=3\times10^8$~M$_\odot$ beginning at $z=10$.
    The effects of a growing potential is mostly easily observed with the early orbit of a low pericentre, high eccentricity dwarf ($r_p=25$~kpc, $\epsilon=0.92$), whose changing orbit cannot be accounted for by dynamical friction alone.}\label{fig:orbs}
\end{figure}

The orbit of the dwarf galaxy was set so that it ended at apogalacticon, with the initial conditions found by integrating backwards without dynamical friction initially, and then with dynamical friction along the previous orbit until convergence of $<5\%$ at all radii was found.
The model was run for dwarf galaxies that at their present day orbits will have a pericentre of between $15$ and $250$~kpc, and a circularity of between $0.2$ and $0.9$.
We assume each point is similar to those around it, and each model was at the centre of a $5$~kpc, $0.05$ circularity box, for those with a pericentre $r_p \le 50$~kpc, a $10$~kpc, $0.1$ circularity box for pericentre $50<r_p\le100$~kpc and a $25$~kpc, $0.1$ circularity box for $100<r_p\le250$~kpc, for a total range of $12.5$--$262.5$~kpc and $0.175$--$0.925$ in circularity.
A sample of these orbits (Fig.~\ref{fig:orbs}) illustrates the effect of varying eccentricity and pericentre in a growing host halo with dynamical friction.

\section{Results}\label{sec:Res}

\begin{figure}
  \includegraphics[width=\textwidth]{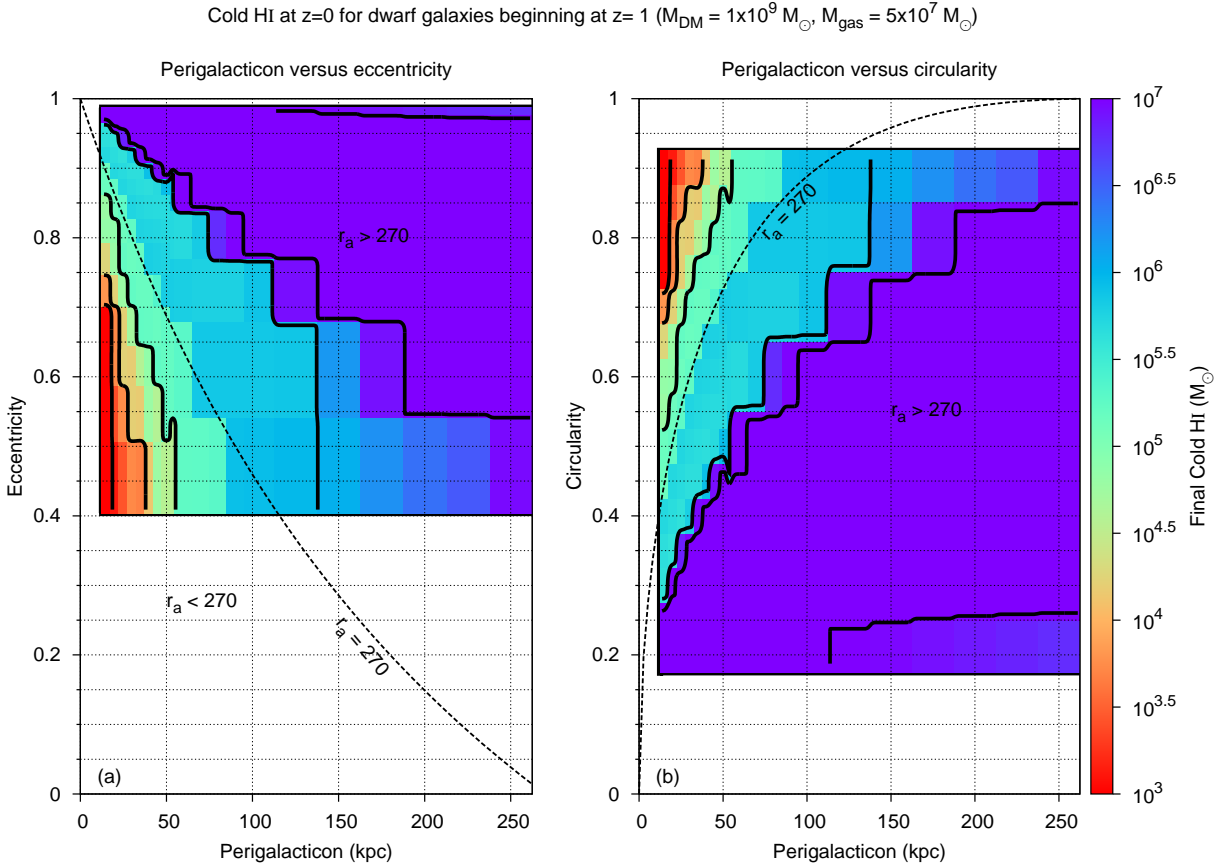}
  \caption{The final neutral hydrogen mass for dwarf galaxies beginning at $z=1$.
  The figures show equivalent information: (a) is the orbit eccentricity vs. the radius of closest approach (perigalacticon) today;
  (b) is the orbit circularity vs. perigalacticon (see \S\ref{ssec:orbit}).
  The rectangular box in both figures shows the range of parameter space explored in this work (see \S\ref{sec:Res}).
  The dashed line corresponds to an apogalacticon radius $r_a=270$~kpc, the radius at which dwarf galaxies suffer strong \HI\ depletion in M31 and the Galaxy.
  The greyscale (color scale in online version) depicts the amount of \HI\ retained by a dwarf galaxy observed today.
  Although the gas in each phase was tracked down to $10$~M$_\sol$ we restrict the range above to $10^{3}$--$10^{7}$~M$_\sol$ for readability.
  Galaxies with $r_a<270$~kpc can only retain $M_{\HIsub}\lesssim{}5\times10^5$~M$_\sol$.}\label{fig:z=1}
\end{figure}

Beginning at $z=1$--$3$ allows most halos to retain a large portion of their cold gas, while beginning at higher redshift $z=10$ means most halos have been stripped of nearly all their cold gas.
In all cases halos lose over $90\%$ of gas mass to warm phase stripping resulting from early heating by internal star formation.
As could be expected, highly circular orbits at low pericentre show the strongest gas loss, with the final mass displayed in Fig.~\ref{fig:z=1}.
This is large gas loss attributable to an on-average larger number of bursts and the increased halo density at low pericentres.
We note that finishing at apogalacticon will always result in later bursts for the same pericentre, but the results remain if the models begin at apogalacticon in which case the bursts will always be earlier.

A large portion of the parameter space exists in which a halo will always be within $270$~kpc and retain over $1\times10^4$~M$_\odot$ of gas.
Numerous other halos will spend a portion of their orbit within this $270$~kpc limit, but as the model finishes at apogalacticon, these subhalos may have experienced less pericentres than would be expected if they were within $270$~kpc today.

\begin{figure}
  \centering
  \includegraphics[width=\textwidth]{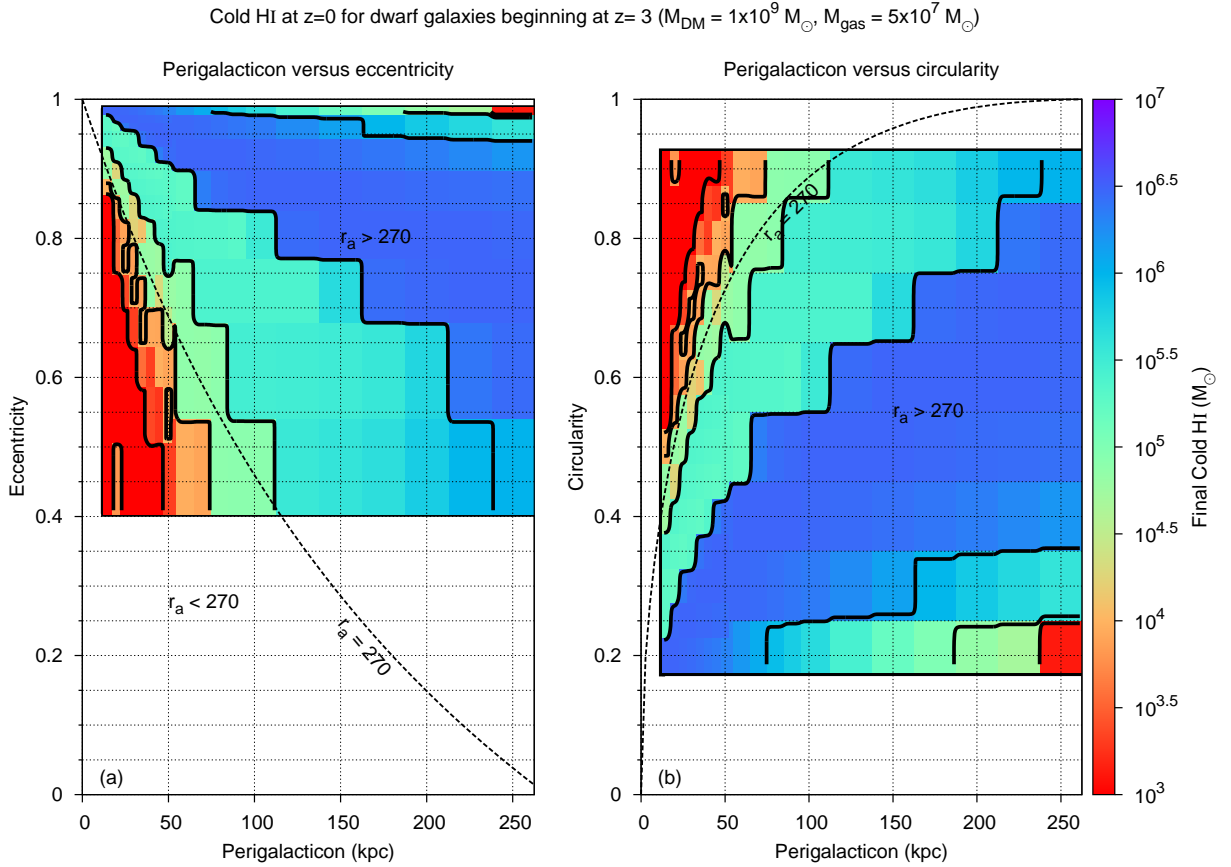}
  \caption{The final neutral hydrogen mass for dwarf galaxies beginning at $z=3$ (Color scale available online).
    See caption to Fig.~\ref{fig:z=1}.
    Galaxies with $r_a<270$~kpc can only retain $M_{\HIsub}\lesssim{}10^5$~M$_\sol$.}\label{fig:z=3}
\end{figure}

For those subhalos which begin at $z=3$ (Fig.~\ref{fig:z=3}) no halos are always within $270$~kpc and maintain over $1\times10^4$~M$_\odot$ of \HI{}; although some halos should be detectable with over $1\times10^3$~M$_\odot$ of gas remaining that will always be within $270$~kpc.

\begin{figure}
  \centering
  \includegraphics[width=\textwidth]{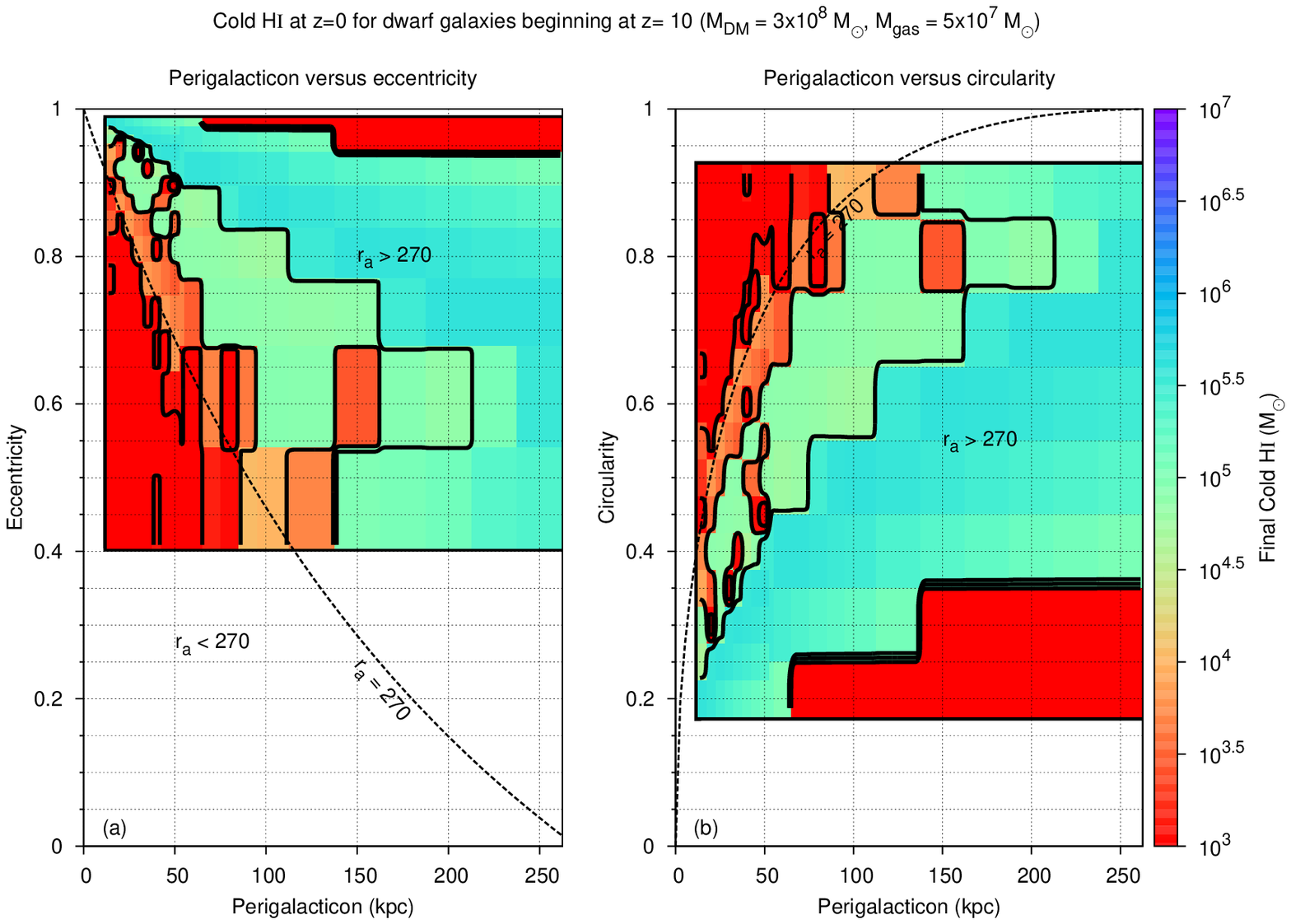}
  \caption{The final neutral hydrogen mass for dwarf galaxies beginning at $z=10$ (Color scale available online).
    See caption to Fig.~\ref{fig:z=1}.
    Galaxies with $r_a<270$~kpc can only retain $M_{\HIsub}\lesssim{}10^3$~M$_\sol$.}\label{fig:z=10}
\end{figure}

The low circularity, high pericentre drop in gas mass is a result of the model finishing at apogalacticon.
This effect is much more pronounced for halos beginning at $z=10$ displayed in Fig.~\ref{fig:z=10}, here the initial velocities may exceed $1000$~km~s$^{-1}$.
The long time in orbit, and lower virial mass of the dwarf galaxies in this run means that nearly all halos that retain gas will not have an apogalacticon within $270$~kpc, that is, any observed halos with gas that began at $z=10$ will only be observed within $270$~kpc for a portion of their orbit.  

\begin{figure}
  \centering
  \includegraphics[angle=-90,width=\textwidth]{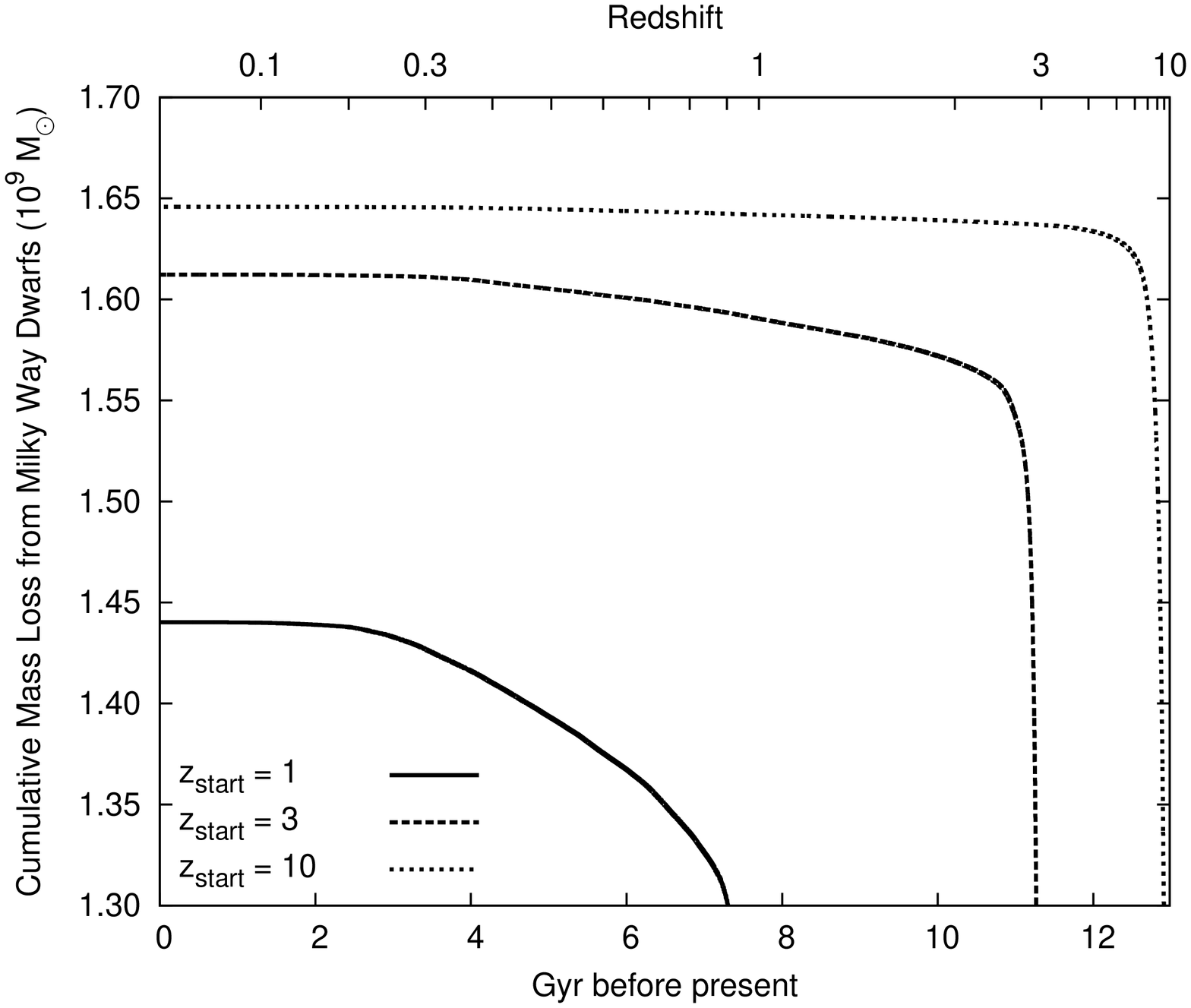}
  \caption{The total dwarf \HI{} loss vs cosmic time due to ram pressure stripping. (Tidal stripping from more massive galaxies, e.g. the Magellanic Stream, is likely to lead to higher mass loss rates at later epochs.) This is calculated by applying the weighting of each dwarf (see \S\ref{sec:Comp}) and multiplying by the number of known dwarf galaxies orbiting the Galaxy ($\sim33$) today. We note that the amount of dwarf galaxies seen today is a small fraction of the total dwarf galaxies that have existed over time, with many destroyed by the Milky Way previously and others unobserved, the total mass loss above will similarly be only a small fraction of the total mass loss from dwarfs over time. The majority of this mass loss occurs very early on as the large amount of cold gas allows rapid star formation to occur, quickly heating and expanding the gas. This low density gas far from the centre of the potential well is easily removed quickly, before a plateau of mass loss is eventually reached.}\label{fig:cummloss}
\end{figure}

\begin{figure}
  \centering
  \includegraphics[angle=-90,width=\textwidth]{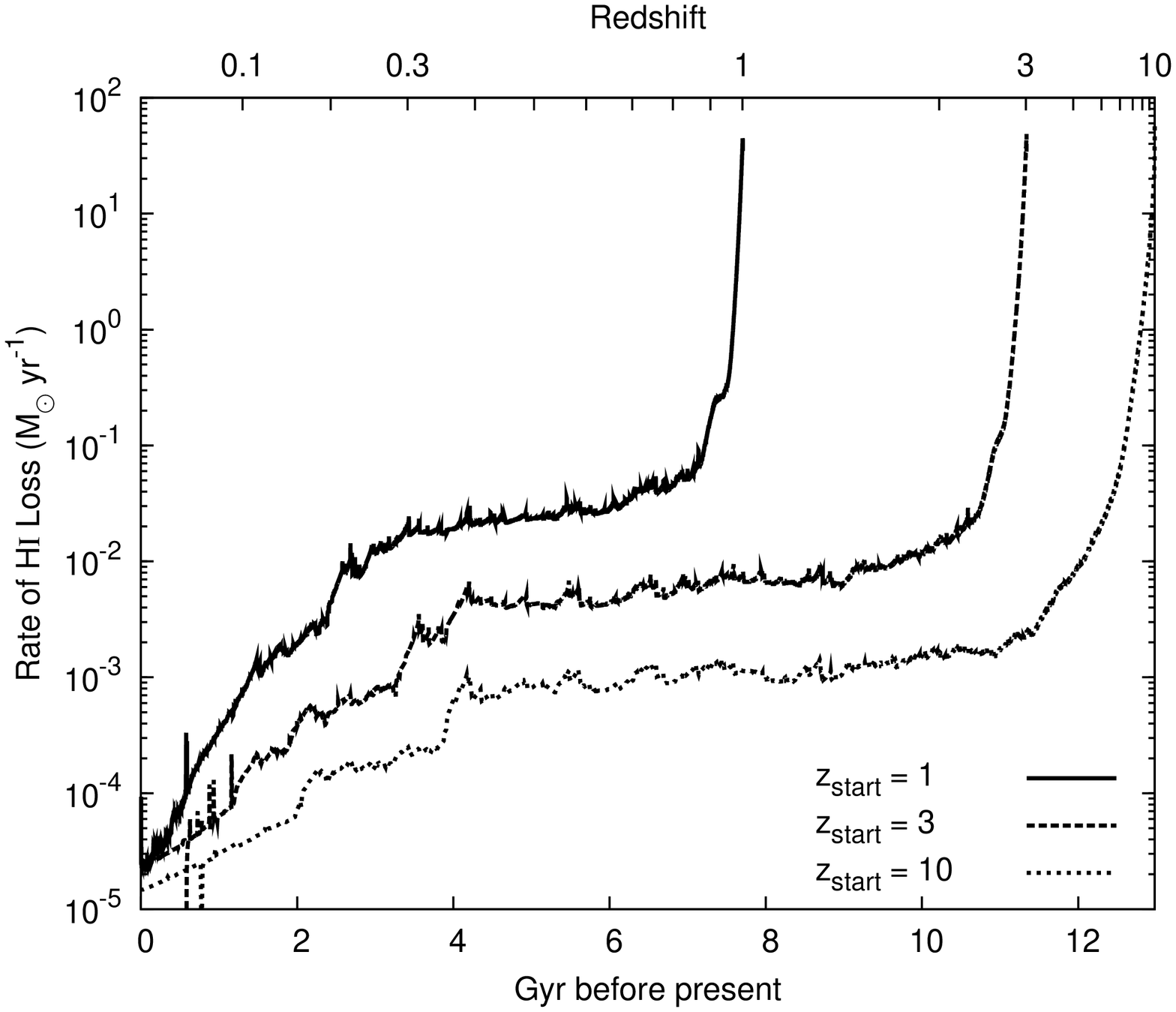}
  \caption{The total dwarf \HI{} loss rate vs cosmic time due to ram pressure stripping. (Tidal stripping from more massive galaxies, e.g. the Magellanic Stream, is likely to lead to higher mass loss rates at later epochs.) This is calculated by applying the weighting of each dwarf (see \S\ref{sec:Comp}) and multiplying by the number of known dwarf galaxies orbiting the Galaxy ($\sim33$) today. We note again that the mass loss rate will only be a small fraction of the actual mass loss rate over all time as it does not include dwarfs that have been destroyed by the Milky Way or are presently unobservable. The period of greatest mass loss occurs when initial star formation heats the \HI{} allowing its removal from the dwarf galaxy before it cools down, a consequence of our model is that this occurs at the same time for all dwarf galaxies exacerbating this mass loss. This time period of greatest star formation rate is also when the supernova-driven winds are most important at removing gas but unlikely to explain \HI{} depletion \citep{Martin1998,JBH2011} and are only accounted for in a simple way (see \S\ref{sec:Conc}). This figure assumes that the dwarfs outside our parameter space experience similar loss rates to those within.}\label{fig:mloss}
\end{figure}

In all models, at early redshifts the high rate of star formation produces large quantities of warm gas which is removed by tidal forces existing on the dwarf galaxy, and warm gas at the edge of the tidal radius is then quickly removed by stripping of the hot halo. This initial period of extremely high mass loss accounts for over $90\%$ of all mass loss as seen in Fig. \ref{fig:cummloss}.
This initial period of gas loss is followed by a steady state, seen more easily in the mass loss rate versus time in Fig. \ref{fig:mloss}, is reached where cold gas is not warmed rapidly, and a small amount of warm gas can survive stripping, with most of the gas being stripped from dwarfs which are approaching their perigalacticon and experiencing bursts of star formation and the increased impact of the Galactic UV field.

At low redshifts, the rate of stripping plummets as the initial conditions require that dwarf galaxies be at their apogalacticon at $z=0$.
Distributing the dwarf galaxies in phase along their orbits today has a minimal effect on the results presented in Figs.~\ref{fig:z=1}--\ref{fig:z=10}.
The steep rate of gas loss for dwarf galaxies beginning at $z=10$ is also a byproduct of our requirement that dwarfs finish at
apogalacticon today. This leads to the excluded region (marked `RAD' in Fig.~\ref{fig:z=10}). The reason is that the dwarf is forced to orbit at
unrealistic velocities to get back to its apogalacticon within an evolving potential well. This artifact has only a minimal
effect on our results.

\section{Comparison with observations and completeness}\label{sec:Comp}
We now determine the fraction of dwarf galaxies retaining gas within $270$~kpc of the Milky Way and M31.
In order to estimate this, the completeness of the parameter space searched needs to be calculated.
We assume that the infalling halos follow a distribution identical to that of \citet{Wetzel2010}, which used a dissipationless {\it N}-body simulation with $\Lambda$CDM cosmology ($\Omega_{\rm m} = 0.25$, $\Omega_\Lambda = 0.75$, $h = 0.72$, $n=0.97$ and $\sigma_8 = 0.8$) with a particle mass of $1.64\times10^8 h^{-1}$~M$_\sol$.
We note that although dwarf galaxy sized halos cannot be represented in this model at $z=0$ we extend the distribution on the basis that for higher masses there exists no dependence on mass. 
For a Milky Way sized halo at $z=0$ these distributions are
\begin{eqnarray}
  \frac{\rm{d}f}{\rm{d}\eta} = 5.0\eta^{1.05}(1-\eta)^{0.75},\\
  \frac{\rm{d}f}{\rm{d}(r_{\rm peri}/r_{\rm vir})} = 3.6\exp\left\{-[3.4(r_{\rm peri}/r_{\rm vir})]^{0.85}\right\}.
\end{eqnarray}

\begin{figure}
  \centering
  \includegraphics[angle=-90,width=\textwidth]{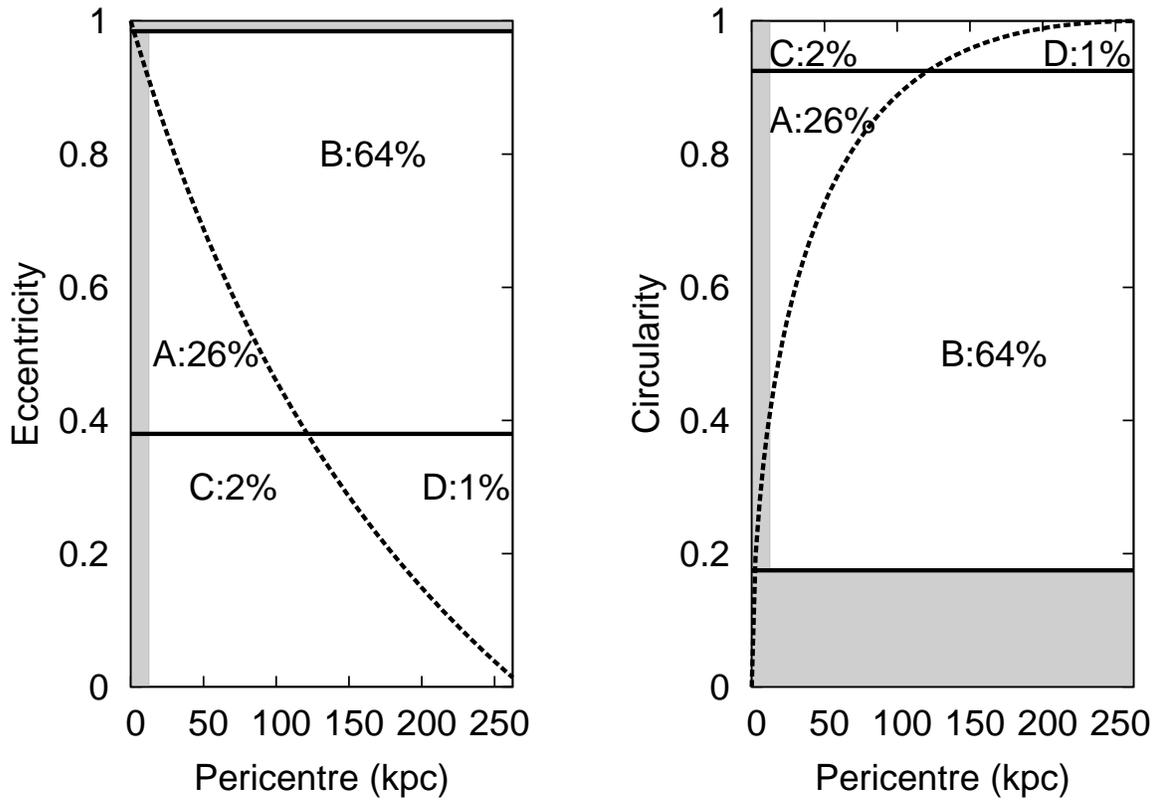}
  \caption{The distribution of subhalos in pericentre-eccentricity (pericentre-circularity) space as a percentage of all subhalos with perigalacticon $r_a<270$~kpc at $z=0$ for a Milky Way sized galaxy, assuming that circularity and pericentre are independent. Calculated using distributions from \citet{Wetzel2010}.}\label{fig:distgraph}
\end{figure}

Taking the circularity and pericentre to be independent, the halos in the parameter space searched will comprise $\sim76\%$ of all halos with a pericentre $<270$~kpc and $\sim70\%$ of all halos, with the distribution within the parameter space shown in Fig.~\ref{fig:distgraph}.
We split the distribution into four main regions: Regions A and B consists of those halos which were within our parameter space and always/sometimes within $270$~kpc respectively.
Regions C and D (halos with circularitys $\eta>0.925$) consists of those halos that were outside our parameter space and always/sometimes within $270$~kpc respectively.
Regions C and D were excluded from the parameter space as the large number of pericentres and low tidal gradients would give an unrealistic amount of bursts as well as the large number of pericentres being computationally expensive.
The shaded region of low circularity/high eccentricity consists of halos which were outside our parameter space and spend so little time within $270$~kpc that they were ignored.
The shaded region of low pericentre is halos that would be stripped due to their very close passages, but the no disk assumption begins to break down as well as the increased computational cost means these halos were not modelled.

In our model all dwarfs end at apogalacticon, however, the results are substantially the same if you begin them at apogalaction (for low redshifts where the orbit does not change significantly over time), we hence assume for analysis that they are at a random point along their orbit, and that this orbit is elliptical, as opposed to the standard Rosetta orbit an Einasto profile will produce.
Under these assumptions, a dwarf galaxy with a pericentre of $r_p<r'$ and eccentricity $\epsilon=\sqrt{1-\eta^2}$ will be within $r'$~kpc with probability
\begin{eqnarray}
  P(r < r') &=& \frac{1}{\pi}\left(\frac{\pi}{2} - \alpha - \arctan\left[\frac{1-(r'/r_p)(1-\epsilon)}{\alpha}\right]\right),\\
  \alpha &=& \sqrt{(r'/r_p -1)(1-\epsilon)(1+\epsilon-(r'/r_p)(1-\epsilon))}. \nonumber
\end{eqnarray}

Within each point in the model we then weight it by its contribution to all halos, with the weight then
\begin{equation}
  w(r') = {\int^{(r_p+x)/r_{\rm vir}}_{(r_p-x)/r_{\rm vir}}\int^{\eta+y}_{\eta-y} P(r<r') \frac{{\rm d}f}{d\eta} \frac{{\rm d}f}{{\rm d}r} {\rm d}\eta{}{\rm d}r},
\end{equation}
where $x$ is half the size of the box in kpc, and $y$ is half the size of the box in circularity.

We assume that all halos with a pericentre $r_p<12.5$~kpc cannot retain gas, and that subhalos with a circularity $\eta>0.925$---halos from regions C and D---will have the same amount of gas, as halos with the same pericentre and a circularity of $\eta=0.9$.
Halos with a circularity $\eta<0.175$ were assumed to contain gas, however, these halos have very eccentric orbits and hence will normally have a near zero weight.

We then calculate what fraction of observed dwarf galaxies contain gas, looking at all galaxies below a given galactocentric radius in $100$~kpc bins out to $2000$~kpc and compare this to the fraction of galaxies---with confirmed detections and excluding the SMC and LMC discussed below---that contain gas from GP09.
We use three cutoffs to distinguish between gas-deficient and gas-rich dwarfs.
A lower cutoff of $10^3$~M$_\sol$ is above the upper-boundary of a number of GP09 galaxies
A mid-range cut-off of $10^{4}$~M$_\sol$ is above the upper-boundary of even more galaxies in the GP09 sample.
Finally an upper cut-off of $10^{5}$~M$_\sol$, this cut-off is below the \HI{} mass of any confirmed gas-rich galaxy in the GP09 sample.
These cut offs do not accurately represent the physical constraints on observations, with the observed being a combination of low cut-offs for nearby Galactic satellites to higher-cutoffs for far-off or M31 satellites.

Under our model, the LMC and SMC being much more massive than the typical dwarf---$M_{\rm LMC}\gtrsim2\times10^{10}$~M$_\sol$ \citep{Schommer1992}, $M_{\rm SMC}\gtrsim3\times10^9$~M$_\sol$ \citep{Harris2006}---will be able to retain significant quantities of warm gas against ram-pressure stripping.
This protection from stripping would allow a significant quantity of gas to survive until the present day, with only tidal interactions---potentially those between the LMC and SMC---removing material from either dwarf galaxy.
Although tidal stripping is included in our model, it is too simplistic to model the multi-body system present in the Magellanic clouds, with hydrodynamical simulations required to simulate the LMC-SMC-Galaxy system \citep{Lin1995,JBH2007,Nidever2008,Besla2010}.

\begin{figure}
  \centering
  \includegraphics[width=\textwidth]{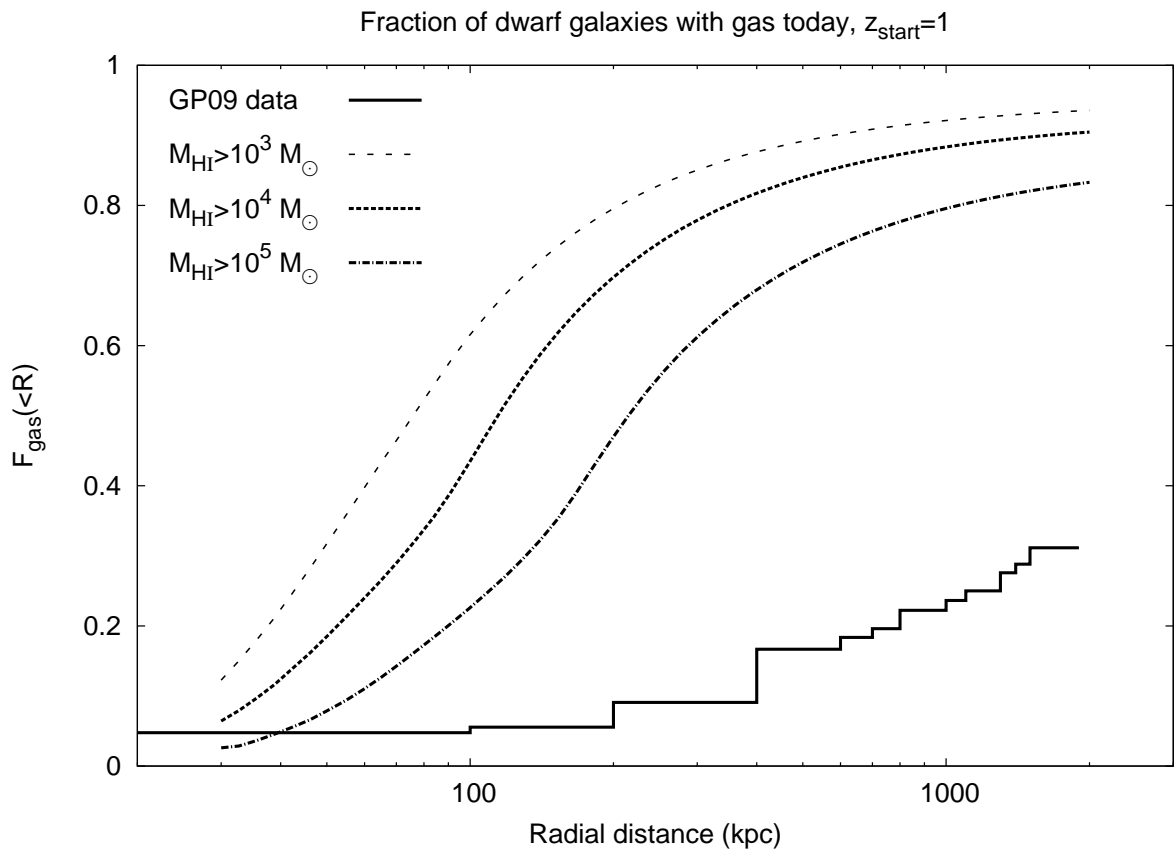}
  \caption{The fraction of dwarf galaxies within a radial distance that contain potentially detectable amounts of \HI{} at $z=0$ versus radius for dwarf galaxies that began at $z=1$.
  We note that the low radius spike in the \citet{Grcevich2009} data (GP09 data) occurs due to a single galaxy---NGC205, a close in satellite of M31.
  Excluding this inner region, all three cutoffs greatly exceed the observed fraction of galaxies that retain gas.}\label{fig:z1results}
\end{figure}

\begin{figure}
  \centering
  \includegraphics[width=\textwidth]{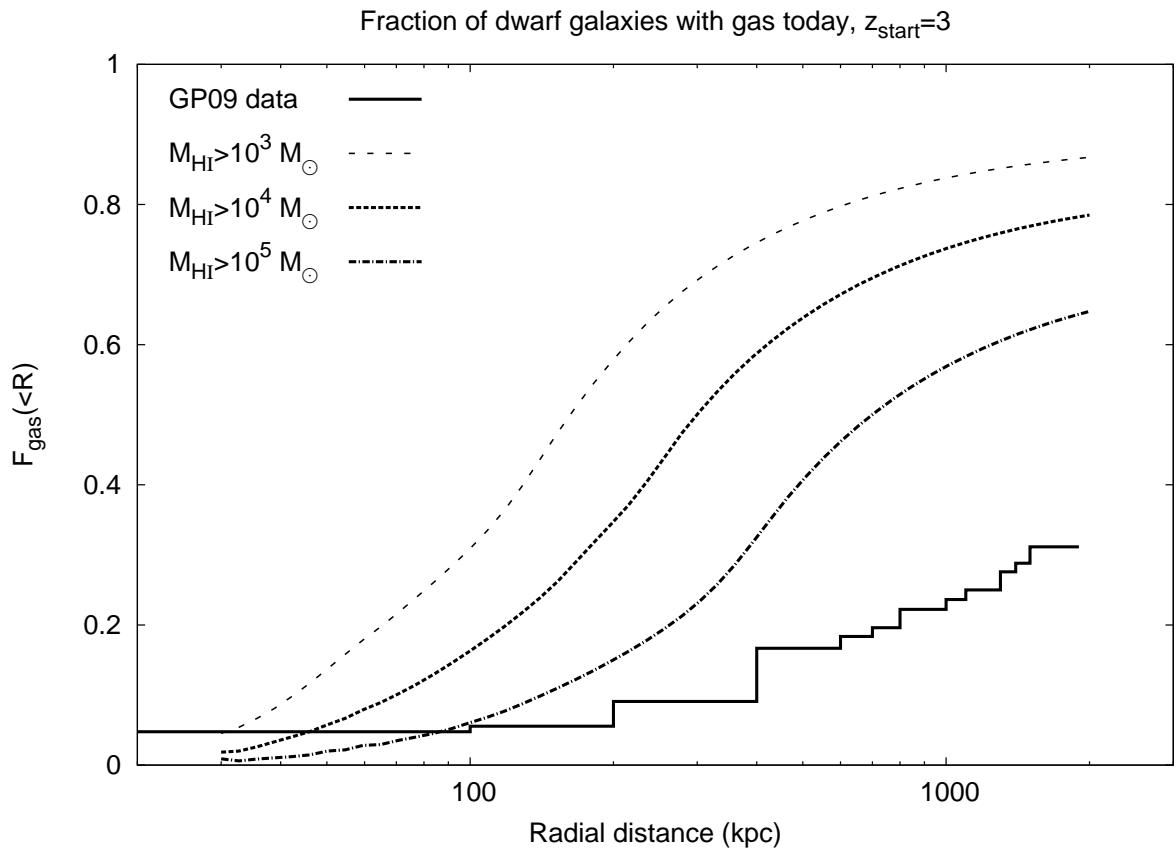}
  \caption{Fraction of dwarf galaxies within a radial distance that contain potentially detectable amounts of \HI{} at $z=0$ versus radius for dwarf galaxies that began at $z=3$.
  Excluding the inner region (see Fig.~\ref{fig:z1results}) the cut-offs of $10^3$~M$_\sol$ and $10^4$~M$_\sol$ exceed the observed fraction of galaxies that retain gas.
The $10^5$~M$_\sol$ cutoff closely follows the cutoff before exceeding it at large radius, this is likely a selection effect (see text).}\label{fig:z3results}
\end{figure}

For models beginning at $z=1$ (Fig.~\ref{fig:z1results}) we see a large over estimation for all gas cut-off masses.
For models beginning at $z=3$ (Fig.~\ref{fig:z3results}) again the predicted fraction exceeds the observed fraction of dwarfs, but predicts a lower fraction than models beginning at $z=1$.
Discovery of new gas-rich dwarfs far from the Galaxy may raise the observed fraction, but this may not explain the over-abundance predicted close in, where gas-rich dwarfs would be more detectable than further gas-deficient dwarfs due to more ongoing star formation. 

\begin{figure}
  \centering
  \includegraphics[width=\textwidth]{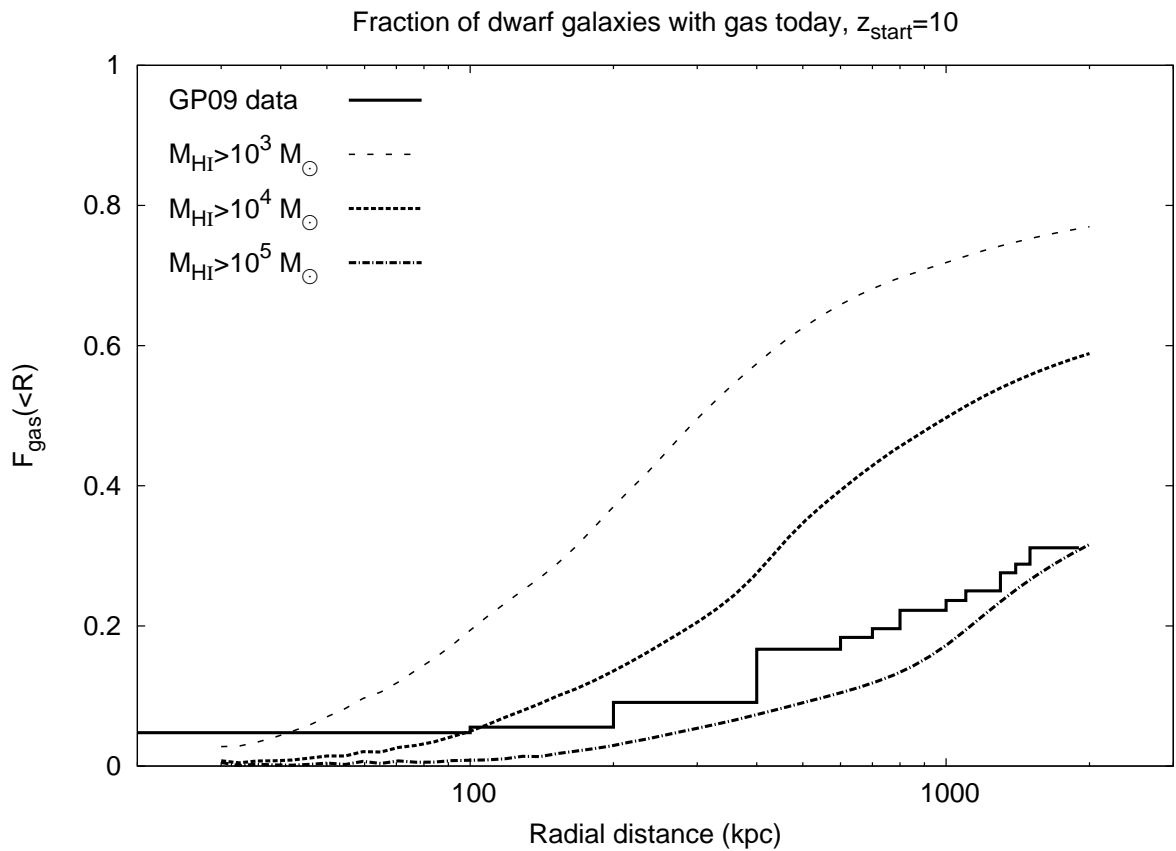}
  \caption{Fraction of dwarf galaxies within a radial distance that contain potentially detectable amounts of \HI{} at $z=0$ versus radius for dwarf galaxies that began at $z=10$.
Again the fraction of close-in dwarfs with gas observed exceeds that of the model (see Fig.~\ref{fig:z1results}), at higher radii are bracketed by the $M_{\HIsub}=10^4$~M$_\sol$ and $M_{\HIsub}=10^5$~M$_\sol$ cutoffs, discussed further in \S\ref{sec:Comp}.}
  \label{fig:z10results}
\end{figure}

For dwarf galaxies that begin their orbits at $z=10$ (Fig.~\ref{fig:z10results}), the $M_{\rm gas}>10^3$~M$_\sol$ and $M_{\rm gas}>10^4$~M$_\sol$ cut-offs again overestimates the fraction of halos which will contain gas.
The $M_{\rm gas}>10^5$~M$_\sol$ cut-off closely follows the shape of the GP09 sample, and is always below this fraction.
At any distance we would expect the \citet{Grcevich2009} fraction to be slightly higher than reality, as the gas poor dwarfs being difficult to see may not be fully accounted for.
That the toy model represents the gas fraction best for halos beginning at $z=10$ adds support to the idea that the Milky Way dwarfs were formed at or before this redshift \citep{Lux2010}.

\section{Summary \& Conclusion}\label{sec:Conc}
This simple toy model reproduces the observed fraction of gas-rich dwarfs, however, there are several factors that were assumed to be negligible which could affect the amount of gas that survives.
The accretion of gas onto dwarfs is unaccounted for in this model.
The prospect of low redshift accretion \citep{Ricotti2009} in particular would greatly increase the chance of a dwarf surviving with gas to the present day, and if included for all dwarfs would likely result in an overestimation of the gas fraction even for dwarfs beginning at late redshifts.
The effects of dust through photoelectric heating and cooling is also ignored.
Due to the low metallicity environments that dwarf galaxies typically have, this effect will be smaller than in larger galaxies, but may still be an important source of heating or cooling.
We also assume that the early extragalactic UV field is uniform in space, a clumpy radiation field around the time of reionization, may have a large impact on the amount of gas that remains cold and protected from ram pressure stripping, this early gas loss could greatly impact the survival around the earliest pericentre passages.
The use of a smooth medium for the gas---compared to a more realistic fractal medium---minimises the cooling of warm gas and allows it to extend beyond the tidal radius.
Even with a large filling factor, the gas will cool much more quickly via metal line cooling than in the smooth medium minimising the large gas loss at the beginning.
Potentially the biggest limitation is the assumption that all of the supernova energy goes into heating the gas to an extremely hot state, much of this energy likely goes into raising cold and warm gas out of the potential well of the dwarf, allowing it to be much more easily stripped.
In particular this will predominantly impact the lowest gas masses, where the smooth gas assumption versus a fractal medium is more likely to have a large impact \citep{JBH2007}.

Due to the large initial gas loss, our star formation rates do not represent that of the majority of dwarfs, with most stars forming early in the dwarfs life, as opposed to a roughly continuous star formation rate with some bursts \citep{Weisz2011}.
This makes our model more suited to explaining dwarfs with these early periods of star formation where a majority of stars are formed, e.g. BK5N, KDG52 \citep{Weisz2011}.

Even with these limitations, we believe that the toy model provides strong support towards internal heating due to early star formation, allowing the gas to more easily stripped due to the larger scale heights and lower density of warm and hot gas.
In particular we were able to reproduce the fraction of dwarf galaxies that retain \HI{} assuming that the dwarf galaxies infall at a redshift of $z=10$ consistent with \citet{Lux2010}.

The covering fraction of warm gas stripped from the dwarfs is expected to be low, with the contrails containing approximately the same amount of warm gas as exists inside the cold streams that cover $\sim2\%$ of the projected area of galaxies \citep{Faucher-Giguere2010}.
This gas breaking away in small clumps of warm material will form a warm component of the hot halo which is stabilised against heat conduction from the halo by cooling \citep{Vieser2007,Vieser2007a} before falling onto the Galaxy as a warm rain.
Even with the majority of dwarf galaxies entering at early redshifts, the large timescale required to complete large pericentre orbits ($13$~Gyr for $r_p=270$~kpc) means that gas from any late infalling dwarfs---such as those from $z=3$ or $z=1$---will only be falling onto the disk of the Galaxy today.
A search for this gas in H$\alpha$ around local galaxies is now being undertaken using the Maryland-Magellan Tunable Filter \citep{Veilleux2010} and the Grantecan Osiris Tunable Filter \citep{Cepa2003} where the expected emission measure of the gas ($\la 0.1$~cm$^{-6}$~pc) may be just large enough to be detectable.

\acknowledgements
M.N. is supported by an Australian Postgraduate Award. J.B.-H. is supported by a Federation Fellowship from the Australian Research Council.
The authors wish to thank Sylvain Veilleux, Ken Freeman, Brent Tully \& Doug Lin for their help and comments throughout this work.

\end{document}